\documentclass[aps,twocolumn,epsf,floats,prl,reprint,nofootinbib,superscriptaddress]{revtex4-1}
\usepackage{graphics,graphicx,epsfig}
\usepackage{amssymb,color}
\usepackage{epsf,epstopdf,wrapfig}
\usepackage{amsmath}
\usepackage{multirow}

\usepackage{color,graphicx}

\usepackage{amsmath}

\begin{document}
\title{Controlling a complex system near its critical point via temporal correlations}

\author{Dante R. Chialvo}
\affiliation{Center for Complex Systems \& Brain Sciences (CEMSC$^3$), Escuela de Ciencia y Tecnolog\'ia. Universidad Nacional de San Mart\'{i}n, San Mart\'{i}n, (1650) Buenos Aires, Argentina}
\affiliation{Consejo Nacional de Investigaciones Cient\'{i}ficas y Tecnol\'{o}gicas (CONICET), Godoy Cruz 2290, Buenos Aires, Argentina}

\author{Sergio A. Cannas}
\affiliation{Consejo Nacional de Investigaciones Cient\'{i}ficas y Tecnol\'{o}gicas (CONICET), Godoy Cruz 2290, Buenos Aires, Argentina}
\affiliation{Instituto de F\'isica Enrique Gaviola (IFEG-CONICET), Facultad de Matem\'atica, Astronom\'ia, F\'isica y Computaci\'on,
  Universidad Nacional de C\'ordoba, (5000) C\'ordoba, Argentina}

\author{Dietmar Plenz}
\affiliation{Section on Critical Brain Dynamics, National Institute of Mental Health, Bethesda, MD (20892), USA}

\author{Tom\'as S. Grigera}
\affiliation{Consejo Nacional de Investigaciones Cient\'{i}ficas y Tecnol\'{o}gicas (CONICET), Godoy Cruz 2290, Buenos Aires, Argentina}
\affiliation{Instituto de F\'isica de L\'iquidos y Sistemas Biol\'ogicos (IFLySiB-CONICET),  Universidad Nacional de La Plata, (1900) La Plata, Buenos Aires, Argentina}
\affiliation{Departamento de F\'isica, Facultad de Ciencias Exactas,  Universidad Nacional de La Plata, Argentina}

\begin{abstract}
A wide variety of complex systems exhibit large fluctuations both in space and time that often can be attributed to the presence of some kind of critical phenomena. Under such critical scenario it is well known that  the properties of the correlation functions in space and time are two sides of the same coin. Here we test wether systems  exhibiting a phase transition could self-tune to its critical point taking advantage of such correlation properties.  We describe results in three models: the 2D Ising ferromagnetic model, the 3D Vicsek flocking model and a small-world neuronal network model. We illustrate how the feedback of the autocorrelation function of the order parameter fluctuations is able to shift the system towards its critical point. Since the results rely on universal properties they are expected to be relevant to a variety of other settings.
 \end{abstract}

\maketitle
 
The last decade has witnessed an escalating interest in
complex biological phenomena at all levels including macroevoluction,
neuroscience at different scales, and molecular biology.  The  observed complexity in nature
is often traced to critical phenomena because it
 resembles the complexity found for critical dynamics in models and theory \cite{bak,mora,beggs,reviewBiophys,chialvo,tang,miguel,flocks}.  More specifically, 
 it seems that many biological systems reach a  ``sweet spot'' where they attain maximal susceptibility, i.e., sensitivity to changes in the environment, while maintaining internal order. 
 
\begin{figure} [ht]
\centerline{
\includegraphics[width=0.4\textwidth]{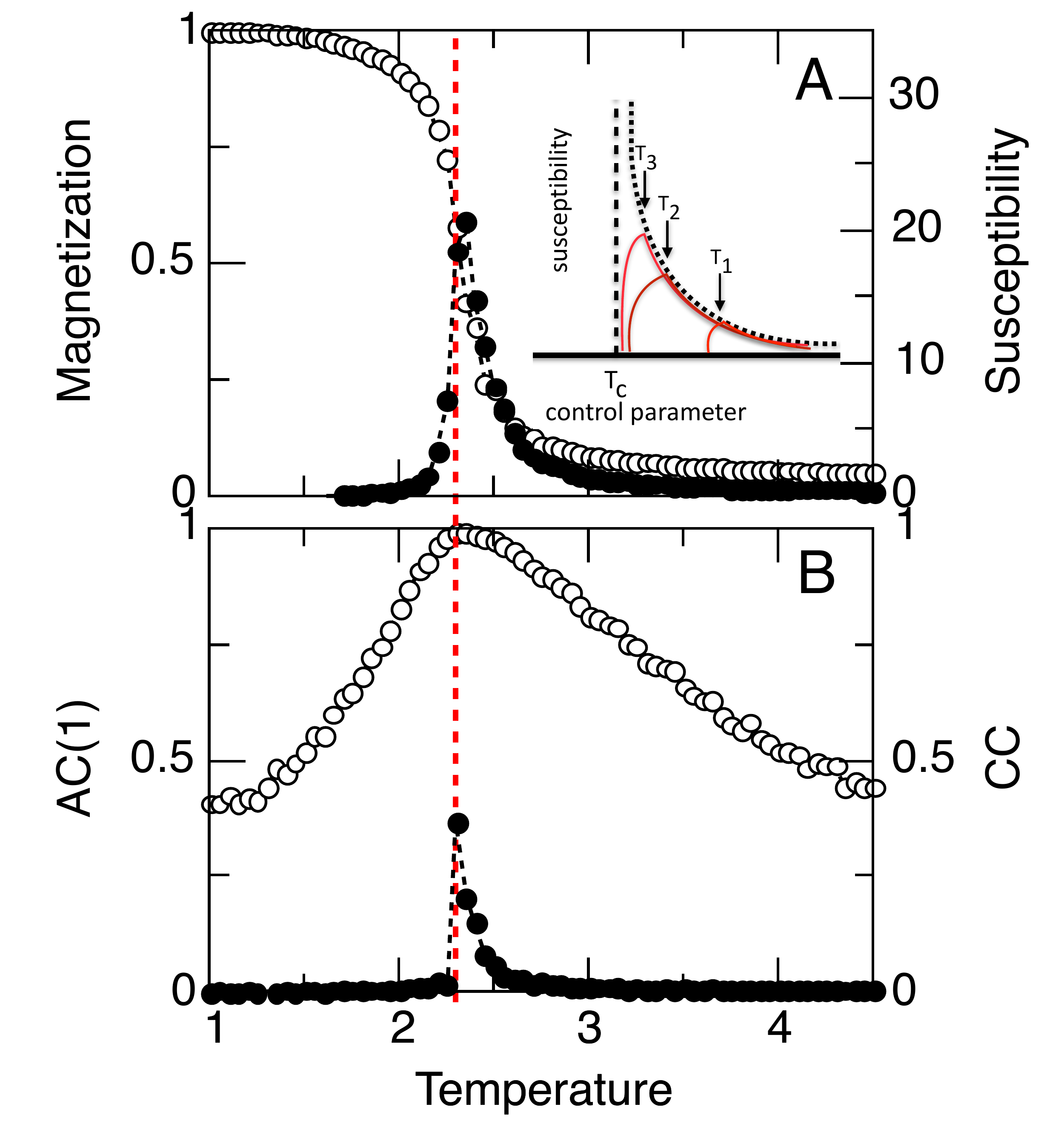}}
\caption{Autocorrelation peaks with susceptibility at $T_c$ in the equilibrium Ising model. Panel A: The order parameter (magnetization; open circles) and susceptibility (filled circles) as function of  temperature $T$.  Panel B: Corresponding average pairwise correlation ($CC$; filled circles) and first autocorrelation coefficient ($AC(1)$; open circles) of the magnetization fluctuations around the instantaneous mean. Dashed vertical line denotes $T_c$. (System size $N=32^2$, $10^4$ MC steps). The inset shows a cartoon of the expected susceptibility as a function of the control parameter for three systems of increasing sizes, where arrows indicate the corresponding optimal points $T_1, T_2,T_3$.  }
\label{fig:Ising-equilibrium}
\end{figure}

 At present  it is not clear how such a critical state can be reached or even
maintained.   For a complex system like the brain, one might imagine that  its control parameters
be hard-wired genetically, selected by a long evolutionary process to
at a critical point that is biologically most advantageous for survival.   However, the critical values of the control parameters
\emph{depend on system size} \cite{DOMB}, and thus for biological systems to take advantage of critical dynamics they would need 
to adjust the control parameter as systems contract or expand. We exemplify this problem in  Fig.~\ref{fig:Ising-equilibrium} (inset)  which sketches how the peak of the susceptibility, the property to be maximized,  shifts as the system get larger.

While some physical systems might be large enough that one can assume they are asymptotically near the thermodynamic limit,
we note that most biological systems are of moderate size, and finite-size effects are in principle to be expected \cite{cavagna2014}.  Hence, if the critical point is the best (or only) functioning state for a given biological system, in order to attain it, Darwinian evolution instead of furnishing a set of specific values for the control parameter 
must allow for a \emph{control mechanism} such that systems can reach and stay close to a critical point.  
For a control mechanism to be  biologically plausible  it should utilize only information that either is completely global or completely local.   Here we explore a possibility for such a mechanism.

We show that the first autocorrelation coefficient $AC(1)$  of the order parameter fluctuations can be used  
 to tune a system to the vicinity of its critical
point.  This is possible because $AC(1)$ peaks at the same point as the
susceptibility, yet does so more smoothly than the susceptibility.

This can be understood from dynamic scaling.  The dynamic scaling form
of the time correlation is \cite{HHREVIEW}
\begin{equation}
  \label{eq:time-scaling-hyp}
  C(k,t) = C_0(k) g\left(\frac{t}{\tau_0(k,\xi)}; k\xi \right),
\end{equation}
where $k$ is the observation wavevector, $\xi$ is the correlation
length, the function $g$ is such that $g(t=0)=1$, i.e.\ $C_0(k)$ is
the static correlation function, and the characteristic time obeys
\begin{equation}
  \label{eq:scaling-tau}
  \tau_0(k,\xi) = k^{-z} \Gamma(k\xi;\xi/L) = \xi^z \Omega(k\xi),
\end{equation}
where $g$, $\Gamma$ and $\Omega$ are unspecified scaling functions and
$z$ is the dynamic scaling exponent. Now

\begin{eqnarray}
  \frac{C(k,t=\delta t)}{C(k,0)} &\approx &1 + \delta t \frac{1}{C_0(k)}  \frac{dC(k,t=0)}{dt} \nonumber \\
  &&= 1  + \delta t\xi^{-z} \Omega^{-1}(k\xi) g'(0;k\xi).
\end{eqnarray}

For a global quantity, $k=0$ (See Suplementary Material) so that
\begin{equation}
  \frac{C(k=0,\delta t)}{C(k=0,t=0)} \sim   1- A\, \xi^{-z} \sim  1- A\,  (T-T_c)^{z \nu},
\end{equation}
where $A>0$ is a time dependent constant. Hence, the normalized time correlation has a maximum at $T=T_c$,  for fixed $\delta
t$.

The main idea is demonstrated here by applying it to three well understood systems, namely the ferromagnetic Ising model,  the Vicsek model of flocking and a typical neuronal small-world network. We remark that the results are general enough to be also expected in many other systems.

\paragraph{Ising model.}  Fig.~\ref{fig:Ising-equilibrium} illustrates the typical behaviour of the 2D ferromagnetic Ising model at increasing temperatures. The system undergoes a second order phase transition at a critical temperature $T_c$, reflected in a steep change in magnetization as well as a sharp peak in susceptibility (Fig.~1A). Equally distinct  changes are also demonstrated for the correlation properties  of the model computed from appropriate system variables (Fig.~1B). A sharp increase in the average pairwise correlations is observed as the system approaches $T_c$, where  the correlation length matches the size of the system. 
The relatively sharp changes in the spatial correlations contrast with the relatively smoother changes in the temporal correlations, as reflected by the first auto-correlation coefficient $AC(1)$  of the magnetization fluctuations around the mean, which  at $T_c$  approaches unity.

Now we asses how to control the Ising model to stay at the vicinity of the susceptibility peak.  According with the discussion in the introduction, we must restrict ourselves to do it using only either local or global information. In that sense, the time correlations evaluated by $AC(1)$ meet such conditions, because it can be  computed from a temporally delayed version of a global average of magnetization.  In turn, magnetization can be assessed  simply by averaging samples of a relatively large number of sites.
%
\begin{figure} [ht!]
\centerline{
\includegraphics[width=0.485\textwidth]{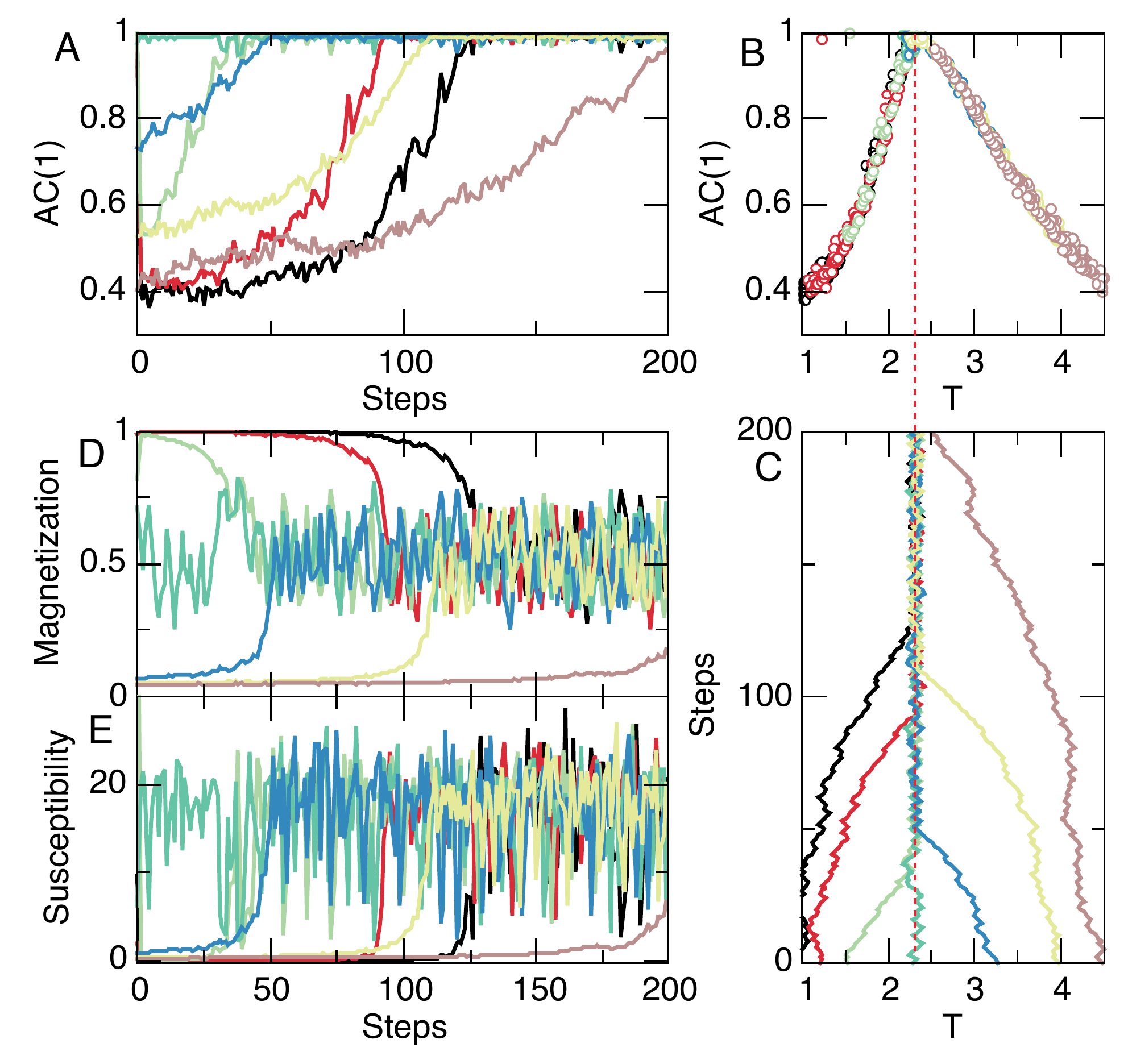}}
\caption{Adaptive control of the Ising model with temperature adjusted iteratively by the autocorrelation of the order parameter.  The data illustrate, for a variety of initial temperatures, the convergence of the system to the vicinity of the expected critical temperature $T_c=2.3$. Panels A,C,D,E show, as a function of iteration steps, the first autocorrelation coefficient of the magnetization' fluctuations, the magnetization, the susceptibility, and the temperature. As seen in Panel C, for any initial temperature the system reaches a state fluctuating near the equilibrium $T_c$ and at the maximum of $AC(1)$ (see panel A). Adaptation parameter $\kappa=0.04$, other parameters as in Fig.~\ref{fig:Ising-equilibrium}.}
\label{Fig2}
\end{figure}

To demonstrate control we proceed  by choosing an initial random temperature and simulate the dynamics for some large number of Montecarlo (MC) steps,  which  we denote as an ``adaptive iteration step'' indexed by $i$. We proceed by estimating the $AC(1)$  of the fluctuations around the mean magnetization during the lapse of time corresponding to the adaptive iteration step $i$ and monitor the change of $AC(1)$ between two consecutive steps $i$, defining
\begin{equation}
 d_i= d_{i-1}\, \text{sign}[AC(1)_{(i)} - AC(1)_{(i-1)}],
 \label{eq:1}
\end{equation}
so that $d$ changes sign when a decrease in $AC(1)$ is detected.  We then use the gradient to its maximum value
\begin{equation}
 \delta_i=(1- AC(1)_{(i)})^2,
 \label{eq:2}
\end{equation}
to change the future temperature $T(i+1)$ according to
\begin{equation}
 T_{(i+1)}=T_{(i)} + \delta * d*\kappa,
 \label{eq:3}
\end{equation}
where $\kappa$ is a small constant that determines how slowly the temperature is adjusted. Its exact values is not crucial for the present results.
Successive iterations of  Eqs.~\ref{eq:1}--\ref{eq:3} demonstrate  convergence of the temperature to the expected value at equilibrium $T_c \sim 2.3$.  Fig.~\ref{Fig2} illustrates typical results for various initial temperatures, which in all cases  converge to the vicinity of $T_c$. We note that  the successive values of the parameters (order, control and $AC(1)$) obtained during the adaptive simulations over-imposes well (i.e., matches) those obtained from equilibrium simulations (i.e.\ the data of Fig.~\ref{fig:Ising-equilibrium}, see Suppl. Material).

\begin{figure} [ht]
\centerline{
\includegraphics[width=0.275\textwidth]{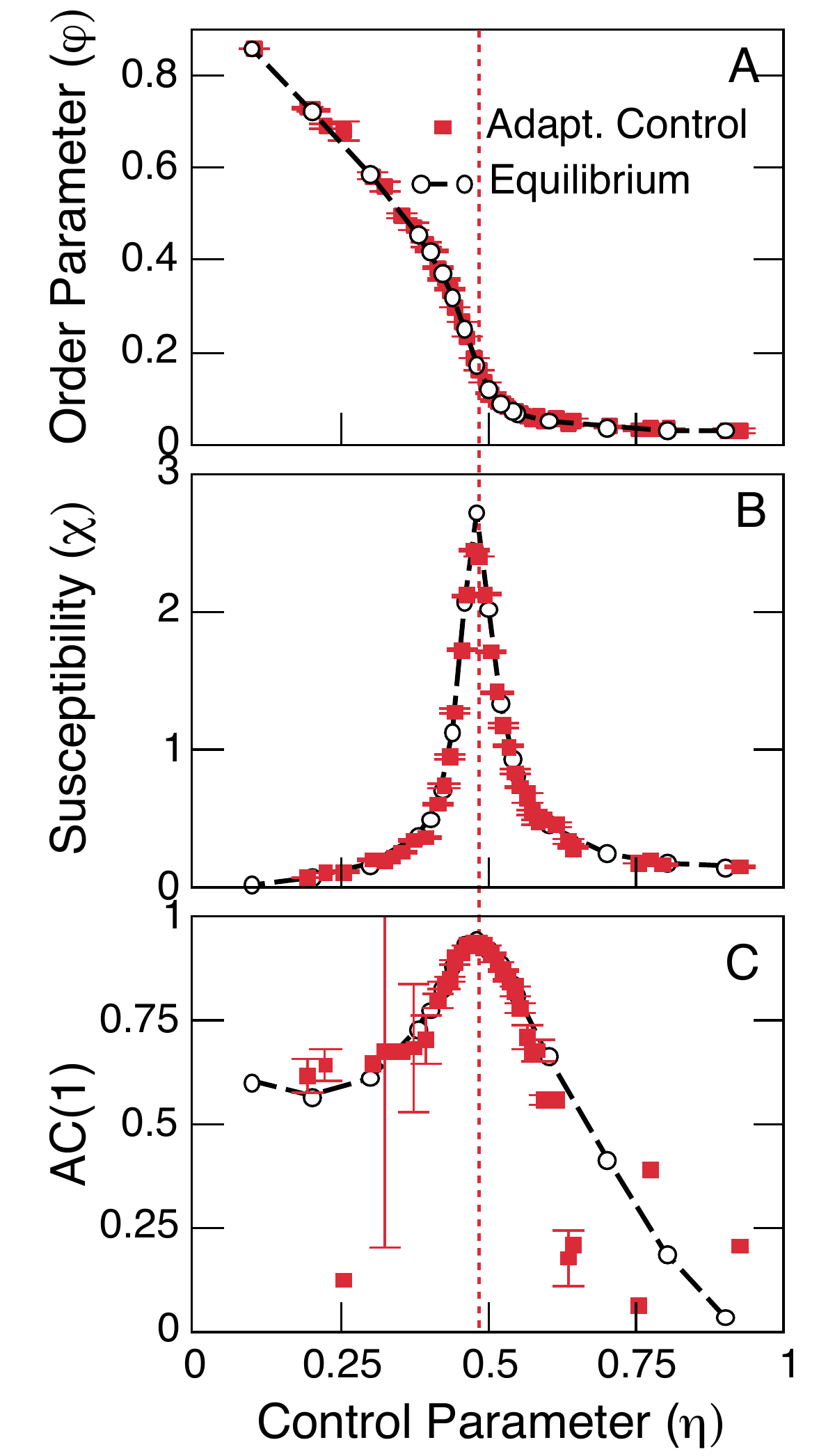}}
\caption{3D Vicsek model at equilibrium and under adaptive control.  Order  parameter $\varphi$ (Panel A), the susceptibility $\chi$ (Panel B) (computed as $\text{var}(\varphi)/N$)  and the first auto-correlation coefficient $AC(1)$ of the polarization fluctuations around the mean (Panel C) as function of $\eta$ the control parameter. Notice the overlap between the equilibrium results (open circles)  and the values reached during the adaptive control (filled squares) for different initial conditions which converge to the critical point denoted by the dashed line  ($\eta_c \sim 0.48$).  $N=1024$, $v_0=1$, $\rho=1.2$.  $\kappa=0.2$ and $10^3$ MC steps per adaptive iteration step.
}
\label{Fig44}
\end{figure}

\begin{figure} [ht]
\centerline{
\includegraphics[width=0.5\textwidth]{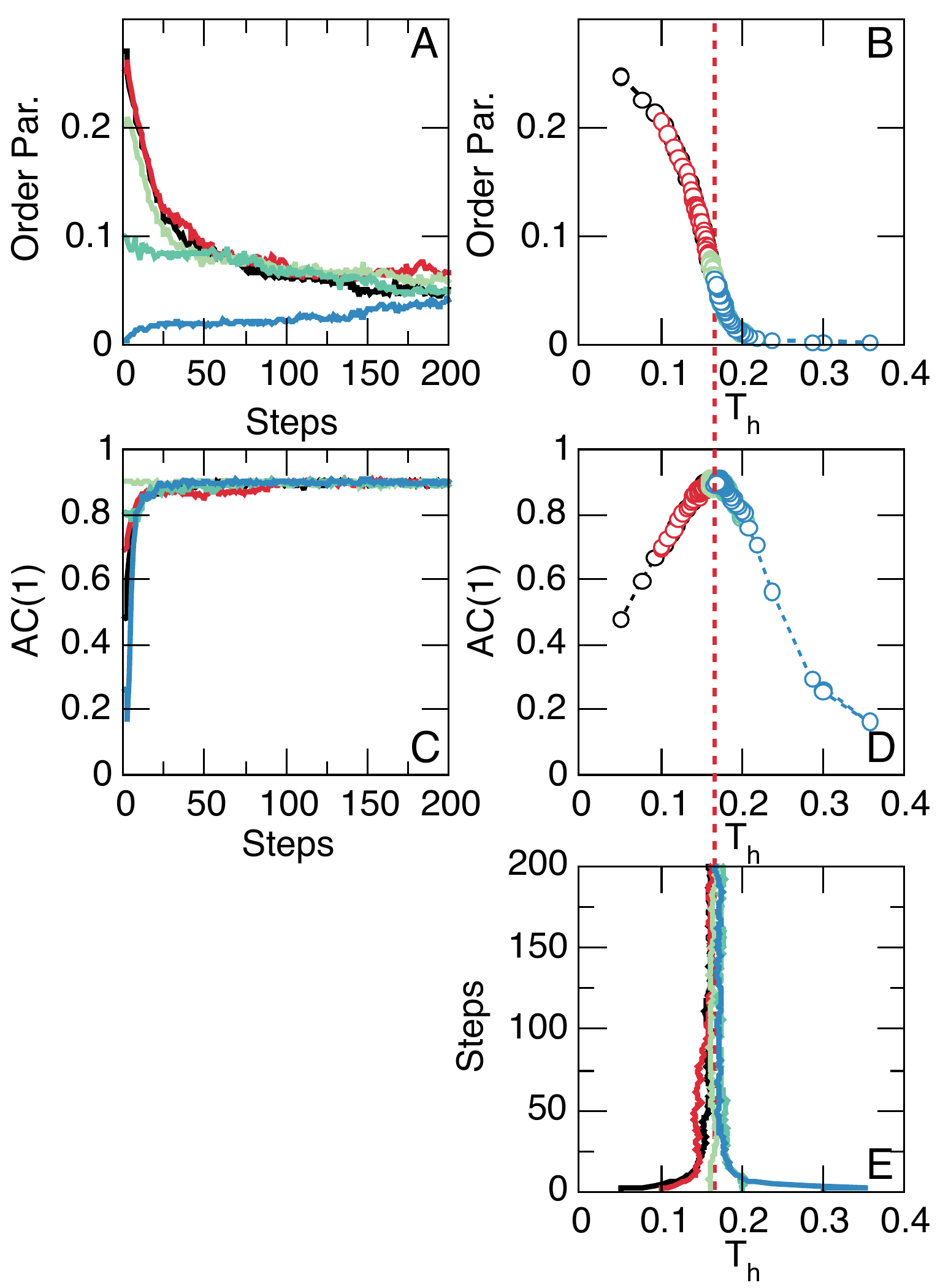}}
\caption{Adaptive control of the neuronal network model.  Data corresponds to numerical solutions of the model starting from five different initial conditions of the control parameter $T_h$. Nodes threshold $T_h$ are adjusted iteratively  according to  Eqs.~\ref{eq:1}--\ref{eq:3}.  Panels A, C and E show the evolution of the order parameter, $AC(1)$ and control parameter $T_h$  respectively. These quantities are plotted against each other in Panels B and D to demonstrate convergence to the critical  value of $T_h$ (dashed red line) and to the maximum of $AC(1)$. $\kappa=0.1$ and $5*10^3$ time steps per adaptive iteration step. Network parameters are: mean degree $<k> = 10$,  rewiring parameter $\beta=0.3$,  system size $N=2^9$. Non null weights chosen from a  distribution $p(w)=\lambda\, e^{-\lambda\, w}$ with $\lambda=12.5$.}
\label{FigNeuro}
\end{figure}

\paragraph{Vicsek model.} We were also able to use the $AC(1)$ function to control  the Vicsek model \cite{VM}, the archetypal model for flocking behavior, towards its critical point.  In this model, $N$ self-propelled particles endowed with a fixed speed $v_0$ move in $d$-dimensional space.  At each time step, positions $\mathbf{r}_i(t)$ and velocities $\mathbf{v}_i(t)$ are updated according to
\begin{align}
  \mathbf{v_i}(t+\Delta t) &= v_0 \mathcal{R}_\eta \left[ \sum_{j\in S_i}
                      \mathbf{v}_j(t) \right], \label{snoiseupd} \\
  \mathbf{r_i}(t+\Delta t) &= \mathbf{r}_i(t) + \Delta t
                             \mathbf{v}_i(t+\Delta t), \label{rFU}
\end{align}
where $S_i$ is a sphere of radius $r_c$ centered at $\mathbf{r}_i(t)$.
The operator $R_\eta$ normalizes its argument and rotates it
randomly within a spherical cone centered at it and spanning a solid
angle $\eta \Omega_d$, where $\Omega_d$ is the area of the unit sphere
in $d$ dimensions ($\Omega_2=2\pi$, $\Omega_3=4\pi$).

The order parameter, which measures the degree of flocking, is the normalized modulus of the average velocity
\cite{VM,reviewVicsek},
\begin{equation}
\varphi \equiv \frac{1}{N v_0} \left| \sum_{i=1}^{N} \mathbf{v}_i \right|.
\label{opVM}
\end{equation}
$\varphi\in[0,1]$, with $\varphi = O(1/\sqrt{N})\sim 0$ in the
disordered phase and $\varphi = O(1)$ in the the ordered phase.
We choose $\Delta t= r_c = 1$, so that the control parameters
are the noise amplitude $\eta$, the speed $v_0$ and the
number density $\rho = N/V$, where $V=L^d$ is the volume of the
(periodic) box.
We apply Eqs.\ref{eq:1}--\ref{eq:3} to this model, using $\eta$ as control parameter and keeping the density fixed.  For comparison we over-plotted  results from equilibrium runs with values taken during adaptive control of the simulations (Fig.~\ref{Fig44}).  The close match demonstrates that the technique is able to control the flock model near its critical noise amplitude, $\eta_c \sim 0.48$ (see further details in Suppl. Material).  

\paragraph{Neuronal network model.} Successful control was further demonstrated for a neural network model \cite{haimovici} consisting of a network of interconnected nodes together with a dynamical rule. The model exhibits  a second order phase transition \cite{Mahdi} on a region of parameters. The model matrix of interactions follows a small-world topology and each node exhibits  discrete state excitable dynamics, following the Greenberg-Hastings model \cite{Greenberg}. Briefly, each node  is assigned one of three states: quiescent $Q$, excited $E$, or refractory $R$,  and the transition rules are: 1) $Q \rightarrow E$  with a small probability $r_1$ ($ \sim 10 ^{-3}$), or if the sum of the connection weights $w_{ij}$ with the active neighbors ($j$) is higher than a threshold $T_h$, i.e., $\sum w_{ij} > T_h$ and $Q  \rightarrow Q$  otherwise; 2) $E \rightarrow R$  always; 3) $R \rightarrow Q$ with a small probability $r_2$ ($\sim 10 ^{-1}$)  delaying the transition from the $R$ to the $Q$ state for some time steps. Parameters $r_1$ and $r_2$, which determine the time scales of self-excitation and of recovery from the excited state, respectively, were kept fixed and  $T_h$ was updated  according to control Eqs. (5--7). The density of  active nodes, i.e. in state  E,  in each time step was taken as the order parameter.  As  shown for the previous models, $AC(1)$ was able to move and maintain  the system near  its critical point (here $T_h\sim0.16$) (Fig. \ref{FigNeuro}).

The present results applies, with some differences, to $1^{st}$ or $2^{nd}$ order phase transitions and also for low dimensional dynamical systems exhibiting continuous or discontinuous bifurcations from fixed points to limit cycles which  can be controlled near the bifurcation point (see Suppl. Material).

In conclusion we have demonstrated, in three paradigmatic cases, how the autocorrelation function of the order parameter fluctuations allows to establish a feedback loop for the control parameter to shift the system towards its critical point.  Our results build on two previous lines of work which come close to describe the control strategy. One is the view of self-organized criticality \cite{bak} as a feedback between order and control parameters \cite{model1}. The other line relates to  forecasting of an upcoming tipping point via the generic slowing down of criticality \cite{tipping,tipping2}. 
The current results go beyond these previous approaches by demonstrating a mechanism that may explain the presence of criticality in some systems, and furthermore providing a  strategy of control amenable of practical implementations in different areas. 
For instance, in neuroscience, this approach could be realized with optogenetical targeting \cite{hollopaper} to clamp cortical networks to any desired dynamical state, helping to predict its influence on perceptual performance.
 
\paragraph{Acknowledgements.} Supported by grant 1U19NS107464-01 from the NIH BRAIN Initiative (USA) and by CONICET (Argentina).


\begin{thebibliography} {50}

\bibitem{bak} P. Bak, \emph{How nature works: The science of self-organized criticality.} Springer Science, New York (1996).

\bibitem{chialvo}D.R. Chialvo, 
 \emph{Nature Physics} {\bf 6,} 744  (2010).

\bibitem{mora}T. Mora \& W. Bialek, 
\emph{J. Stat. Phys.}  {\bf 144,} 268 (2011). 

\bibitem{reviewBiophys} A.R. Honerkamp-Smith, S.L.  Veatch, S.L. Keller, 
\emph{Biochim. Biophys. Acta} {\bf 1788}, 53 (2009).

\bibitem{beggs} J.M. Beggs  \& D. Plenz, 
 \emph{Journal of Neuroscience}  {\bf 23}, 11167 (2003).

\bibitem{tang}Q.Y. Tang, Y.Y. Zhang, J. Wang, W. Wang, D.R.  Chialvo, 
\emph{Phys. Rev. Lett.}  {\bf 118,} 088102 (2017).


 \bibitem{flocks} A. Cavagna, I. Giardina, T.S. Grigera, 
 \emph{Physics Reports} {\bf 728,}1--62 (2018).

\bibitem{miguel} M.A. Mu\~noz, 
\emph{Rev. Mod. Phys.} {\bf 90,} 031001(2018).

\bibitem{DOMB} M.N. Barber, Finite-size scaling, in: C.~Domb and J.~L.~Lebowitz (eds.), \emph{Phase transitions and critical phenomena}, Academic Press, London (1983).

\bibitem{cavagna2014} A. Attanasi,  et al.
\emph{Phys. Rev. Lett.} {\bf 113,} 238102 (2014).

\bibitem{HHREVIEW} P.C. Hohenberg \& B.I. Halperin, \emph{Rev.\ Mod.\ Phys.} {\bf 49,} 435 (1977).

\bibitem{VM} T.  Vicsek, et al.
\emph{Phys. Rev. Lett.} {\bf 75,} 1226 (1995).

\bibitem{reviewVicsek}T. Vicsek \& A. Zafeiris, \emph{Physics Reports}  {\bf 517,} 71 (2012).

 \bibitem{haimovici}A. Haimovici, E. Tagliazucchi, P. Balenzuela, D.R. Chialvo, 
 \emph{Phys. Rev. Lett.} {\bf 110,} 178101 (2012).
 
\bibitem{Greenberg} J.M. Greenberg \& S.P. Hastings, \emph{SIAM (Soc. Ind. Appl. Math.) J. Appl. Math.} {\bf 34,} 515, (1978).

\bibitem{Mahdi}  M. Zarepour, J.I. Perotti,  O.V. Billoni,  D.R. Chialvo  S.A. Cannas. https://arxiv.org/abs/1905.05280 (2019).



\bibitem{model1} D. Sornette, A. Johansen, I. Dornic,  
\emph{J. Phys. I} {\bf 5,} 325--335 (1995).
 \bibitem{tipping} M. Scheffer, J. Bascompte, W.A. Brock, et al,
 \emph{Nature} {\bf 461} 53--59 (2009). 
 \bibitem{tipping2} C. Boettiger, N. Ross, A. Hastings,
 \emph{Theor Ecol} ; {\bf 6} 255--264 (2013).

\bibitem{hollopaper} T. Bellay, A. Klaus, S. Seshadri, D. Plenz, \emph{eLife} {\bf 4} e07224 (2015).

\end{thebibliography}
\end{document}